\documentclass[aps,superscriptaddress,twocolumn]{revtex4}

\usepackage{graphicx}
\usepackage{amssymb,amsmath,amsfonts}
\usepackage[mathscr]{eucal}
\usepackage{color}
\usepackage[breaklinks=true,colorlinks=true,linkcolor=blue,urlcolor=blue,
citecolor=blue]{hyperref}

\usepackage{physics}

\begin{document}

\title{First-passage times in conical varying-width channels biased by a
transverse gravitational force: Comparison of analytical and numerical
results}

\author{Ivan Pompa-Garc\'ia}
\affiliation{Physics Department, Universidad Aut\'onoma Metropolitana-Iztapalapa,
San Rafael Atlixco 186, Ciudad de M\'exico, 09340, M\'exico}
\author{Rodrigo Castilla}
\affiliation{Engineering Faculty, Universidad Nacional Aut\'onoma de M\'exico,
Ciudad Universitaria, Ciudad de M\'exico, 04510, M\'exico}
\author{Ralf Metzler}
\affiliation{Institute of Physics and Astronomy, University of Potsdam,
D-14476 Potsdam-Golm, Germany}
\affiliation{Asia Pacific Center for Theoretical Physics, Pohang 37673,
Republic of Korea}
\author{Leonardo Dagdug}
\email{dll@xanum.uam.mx}
\affiliation{Physics Department, Universidad Aut\'onoma Metropolitana-Iztapalapa,
San Rafael Atlixco 186, Ciudad de M\'exico, 09340, M\'exico}

\date{\today}

\begin{abstract}
We study the crossing time statistic of diffusing point particles between the
two ends of expanding and narrowing two-dimensional conical channels under a
transverse external gravitational field.  The theoretical expression for the
mean first-passage time for such a system is derived under the assumption
that the axial diffusion in a two-dimensional channel of smoothly varying
geometry can be approximately described as a one-dimensional diffusion in an
entropic potential with position-dependent effective diffusivity in terms of
the modified Fick-Jacobs equation. We analyze the channel crossing dynamics
in terms of the mean first-passage time, combining our analytical results
with extensive two-dimensional Brownian dynamics simulations, allowing us to
find the range of applicability of the one-dimensional approximation. We find
that the effective particle diffusivity decreases with increasing amplitude
of the external potential. Remarkably, the mean first-passage time for
crossing the channel is shown to assume a minimum at finite values of the
potential amplitude.
\end{abstract}

\maketitle

\section{Introduction}
\label{introduction}

The Brownian motion of molecules, particles, or even living microorganisms
in confined geometries such as pores and channels, plays a key role on
various scales in both nature and technology
\cite{Jacobs,Zw,BDB2010,BDB2015,VBDB2016,RR,KPre,MSP,CCD,SD,Ao,Redner,
BDB2017,Hanggi,Burada,KP,Br,DP,Bauer,Pompa,KPG,Feng,aljaz,KPe09,KPe11}.
Transport in confined geometries within quasi-one-dimensional systems
exhibits a very rich and striking phenomenology and has been studied
in-depth in many contexts, examples including diffusion in human
metabolism, breathing, or medical drug delivery \cite{DD} as well as
in living cells \cite{Brangwynne}, the motion of viruses and bacteria
\cite{Balluffi,JMR}, solid-state and protein nanopores as single-molecule
biosensors for the detection and structural analysis of individual molecules
\cite{Dekker,Keyser,Pedone}, transport in zeolites \cite{Haul}, synthetic
nanopores \cite{Gershowand,Sexton,Kosinska,Berezhkovskii}, microfluidic
devices \cite{Han}, channels in biological systems \cite{Hille}, and
artificial pores in thin solid films \cite{Howorka}.

A universal description of an unbiased Brownian particle is given by the
free diffusion coefficient $D_0$ in homogeneous systems. In heterogeneous
environments with finite characteristic length scales of the disorder,
the particle motion becomes Brownian at times sufficiently exceeding the
correlation time of the system \cite{degennes}, albeit the crossover time
may be significantly delayed \cite{igorg}. When the diffusion takes place
in systems decorated with excluded-volume obstacles, the diffusion may
be locally free and characterized by the diffusivity $D_0$, while at long
times the particle motion is again Brownian but with an effective diffusion
coefficient $D_{\mathrm{eff}}$ \cite{yael,felix,surya,ilpo}.  Typically the
mean squared displacement (MSD) in such systems monotonically crosses over
from the short time behavior $\langle \Delta x^2(t)\rangle\simeq D_0t$
to $\simeq D_{\mathrm{eff}}t$, where $D_0>D_{\mathrm{eff}}$. Another relevant
case is that of confinement by boundaries, in channels or porous media, in
which a significant slowdown of the MSD is effected \cite{DMBZB,BDB,dan}.
Spatial confinement modifies the equilibrium of the system and its dynamical
properties, increasing the hydrodynamic drag on such components and limiting
the configuration space accessible to its diffusing parts \cite{Deen}. In
this sense, asymmetry plays a major role in the transport of a Brownian
particle through a channel \cite{Burada10,Marchesoni09}, Brownian pumps
\cite{Ai06,Ai08}, and Brownian ratchets \cite{Matthias03n,Hanggi}.

In simpler systems transitions across entropic or energetic barriers effect
single-exponential kinetics of processes such as channel-facilitated transport
of solutes to isomerization reactions. Recent experiments with single
biological nanopores, pulling proteins and nucleic acids, as well as
single-molecule fluorescence spectroscopy have raised a number of questions
that stimulated the theoretical and computational investigation of
barrier-crossing dynamics
\cite{Hummer04,Best05,Berezh06,Vanden06,Sega07,Chung09,Pirchi11,Chung12,
Neupane12,Piana12,Zhang12,Chung13,Frederickx14,Gurnev14,Chung15,Truex15,
Best16,Neupane16}.
The quantity of interest in such studies is the time required for the
system to pass over the barrier region. This time, called direct-transit
time or transition path, is a random variable, characterized by the
associated probability density and mean value.

While the transition path quantifies exclusively successful crossing events,
the first-passage time counts the entire time elapsed until the first
accomplished crossing event. First-passage problems arise in an extensive
range of stochastic processes of practical interest \cite{Redner}. Indeed,
examples for first encounter-controlled events \cite{Benichou08} include
chemical and biochemical reactions \cite{Talkner90,Coppey04,aljaz1,denis},
distance-effects on rapid search of signaling molecules \cite{leonid,otto},
trafficking receptors on biological membranes \cite{Holcman04}, animal foraging
\cite{Benichou05,vladpnas,vlapnjp}, and the spreading of sexually transmitted
diseases in a human social network or of viruses across the World Wide
Web \cite{Gallos070}. It is worth mentioning that it was recently shown
that in confining channel geometries the channel-intrinsic resistance is
directly proportional to the mean first-passage times (MFPTs) of the molecule
between the two channel ends \cite{BB22}. Moreover, first-exit times were
analyzed recently in periodic channels with narrow passageways and
position-dependent diffusivity \cite{yongge}.

\begin{figure}
\includegraphics[width=0.3\textheight]{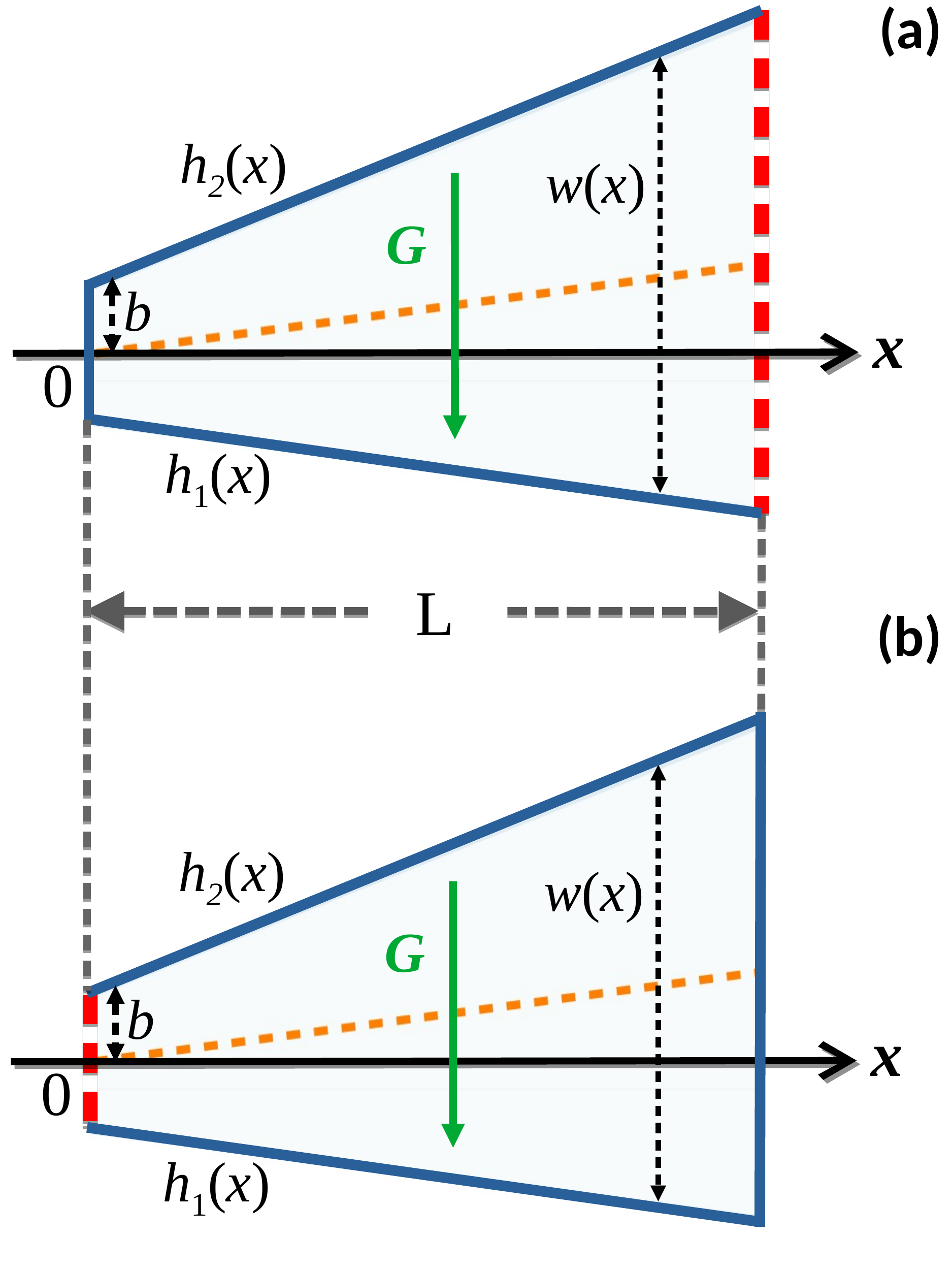}
\caption{Schematic representation of a two-dimensional asymmetric conical
channel formed by straight walls in the presence of a constant transverse
("gravitational") force $G$ (shown as a green downwards arrow). The lower
boundary is given by $h_1(x)=\lambda_1x-b$ (shown as the blue solid line),
while the upper boundary is given by $h_2(x)=\lambda_2x+b$ (shown as the
blue solid line). The channel's variable width is given by $w(x)=h_2(x)
-h_1(x)$, and its straight midline by $y_0(x)=[h_1(x)+h_2(x)]/2$ (shown
as the dotted orange line). Panel (a) shows an expanding channel, i.e.,
the transition of the particle occurs from the narrow to the wide end
($n\rightarrow w$). In such a case, the Brownian particle starts from the
reflecting boundary located at $x=0$ and is then removed by the absorbing
boundary at $x=L$ (shown as the red vertical dashed line). Panel (b) shows
a narrowing channel, i.e., the transition of the particle is from the wide
to the narrow end ($w\rightarrow n$). In this case, the Brownian particle
starts from the reflecting boundary located at $x=L$ and is removed by
the absorbing boundary at $x=0$ (shown as the red vertical dashed line).}
\label{tube1}
\end{figure}

The aim of this article is to study the transitions of diffusing point
particles between the two ends of geometrically expanding and narrowing
two-dimensional conical channels under a transverse gravitational external
field by means of the MFPT, as sketched in Fig.~\ref{tube1}. The theoretical
expression for the mean-fist passage time for such a system is derived
assuming that the axial diffusion in a two-dimensional channel of smoothly
varying geometry can be approximately described as one-dimensional diffusion
with a position dependent diffusion coefficient.

The remainder of this article is organized as follows. After a brief
introduction into the theory of diffusion in geometric channels in Section
\ref{entropic}, in Section \ref{General results} we compare our theoretical
results derived in the framework of the one-dimensional description with the
results obtained from two-dimensional Brownian dynamics simulations. In the
last Section \ref{Conclusions} we discuss our findings and their implications.

\section{Entropic potential and effective one-dimensional description}
\label{entropic}

When diffusion occurs in quasi-one-dimensional channel structures,
one approach to treat the problem is to map the particle motion onto an
effective one-dimensional (1D) description in terms of the diffusion along
the midline $y_0(x)$ of the channel, as a function of the longitudinal
coordinate $x$. The key point of the derivation is the assumption of
equilibration in the transverse direction. The so-called Fick-Jacobs (FJ)
approach consists of eliminating these fast transverse stochastic degrees
of freedom.  The associated, approximate description relies on the modified
Fick-Jacobs-like equation derived by Zwanzig (Zw) for the probability density
$c(x,t)$ in the channel \cite{Zw},
\begin{equation}
\frac{\partial c(x,t)}{\partial t}=\frac{\partial}{\partial x}\left\{D(x)
w(x)\frac{\partial}{\partial x}\left[\frac{c(x,t)}{w(x)}\right]\right\},
\label{FJZw}
\end{equation}
where $w(x)$ is the channel width and $D(x)$ is the position-dependent diffusion
coefficient. Equation (\ref{FJZw}) with a position-independent diffusion
coefficient, $D(x)=D_0$, is known as the ordinary FJ equation
\cite{Fick,Jacobs}. The effective 1D probability density $c(x,t)$ is related
to the 2D probability density $\rho(x,y,t)$ by the projection
\begin{equation}
c(x,t)=\int_{w(x)}\rho(x,y,t)dy.
\label{pd}
\end{equation}

It is well known that confinement in higher dimensions may give rise to an
effective entropic potential in reduced dimensions. In fact, Equation
(\ref{FJZw}) is formally equivalent to the Smoluchowski equation
\begin{equation}
\frac{\partial p(x,t)}{\partial t}=\frac{\partial}{\partial x}\left\{D(x)
e^{-\beta U(x)}\frac{\partial}{\partial x}\left[e^{\beta U(x)} p(x,t)
\right]\right\},
\label{Sm}
\end{equation}
where the entropic potential is given by $-\beta U(x)=\ln[w(x)/w(x_0)]$
with $\beta=1/(k_BT)$, where $k_B$ is the Boltzmann constant, $T$ is the
absolute temperature, and $U(x)$ at $x=x_0$ is taken to be zero.

The expression for the position-dependent effective diffusion coefficient
for a narrow 2D channel of varying width that has a straight midline
suggested by Reguera and Rubi (RR) based on heuristic arguments reads
\begin{equation}
D(x)\approx D_{RR}(x)=\frac{D_0}{\left[1+\frac{1}{4}w'^2(x)\right]^{\eta}},
\label{eq:RR2D}
\end{equation}
where $w'(x)=dw(x)/dx$. This last equation is a generalization of Zwanzig's
expression \cite{Zw}. Alternative derivations of this equation were given
by Kalinay and Percus \cite{KP}, Martens \textit{et al}. \cite{MSP} and
Garc\'ia-Chung  and co-workers \cite{CCD}.

Reguera and Rubi \cite{RR}, Kalinay \cite{KPG}, and later Pompa-Garc\'ia
and Dagdug \cite{Pompa}, studied how Equation (\ref{FJZw}) is modified when
a gravitational and entropic potential coexist. Pompa-Garc\'ia and
Dagdug showed that for an overdamped Brownian particle diffusing in a 2D
asymmetric channel of varying cross-section in the presence of a
constant force in the transverse direction, Equation (\ref{Sm}) takes on
the modified form
\begin{equation}
\frac{\partial c(x,t)}{\partial t}=\frac{\partial}{\partial x}\left\{D(x)
A(x)\frac{\partial}{\partial x}\left[\frac{c(x,t)}{A(x)}\right]\right\},
\label{FJZwG}
\end{equation}
where
\begin{equation}
A(x)=\int_{h_1(x)}^{h_2(x)}e^{-gy}dy=\frac{1}{g}\left[e^{-gh_1(x)}-e^{-g
h_2(x)}\right].
\label{Asym}
\end{equation}
Equation (\ref{FJZwG}) is obtained when a 2D asymmetric channel is bounded
by perfectly reflecting walls given by smooth functions $h_1(x)$ and $h_2(
x)$, and $g=G/k_B T$, where $G$ stands for the constant transverse-direction
force, and $-\beta U(x)=\ln{[A(x)/A(x_0)]}$. In Equation (\ref{Asym}) the
coupling between entropy and energy barriers is reflected by the presence
of the term $\exp\left[gh_i(x)\right]$ ($i=1,2$). For a symmetrical channel,
when $h_1(x)=-h_2(x)$, $A(x)$ reduces to $(2/g)\sinh\left[gh_2(x)\right]$
\cite{KPG}.

Using the projection method, neglecting the second- and higher-order
derivatives of $h(x)$, and setting the parameter $\epsilon=D_x/D_y$,
which expresses the ratio of the diffusion constant in the longitudinal
and transverse direction, equal to infinity (inspired by RR's and Kalinay's
work) Pompa-Garc\'ia and Dagdug found the effective diffusivity with the
same structure as Equation (\ref{eq:RR2D}),
\begin{align}
\begin{split}
\eta_{a}(gw,y_0')&=\frac{1}{\sinh^2\left(\frac{1}{2}gw\right)}\\
&\times\left\{1+\cosh^2\left(\frac{1}{2}gw\right)-gw\coth\left(
\frac{1}{2}gw\right)\right\}\\
&+4\frac{y'_0}{w^{'2}}\left\{y'_0-w'\coth\left(\frac{1}{2}gw\right)\right.\\
&\hspace{3em}\left.+\frac{1}{2}gww'\csch^2\left(\frac{1}{2}gw\right)\right\},
\end{split}
\label{eq:PD3}
\end{align}
where $y_0(x)$ is the midline of the channel. One of the most important
features of Equation (\ref{eq:RR2D}) when $\eta$ is given by Equation
(\ref{eq:PD3}) is the breaking of symmetry for a strong field. Thus while
the predicted diffusivity is $D_0/\left[1 + h_1'^2(x)\right]$ when $G$
tends to infinity, for $G$ going to minus infinity, the predicted diffusivity
is $D_0/\left[1+h_2'^2(x)\right]$. In both cases, this constrains the
Brownian dynamics to one dimension over the boundaries $h_1$ or $h_2$,
respectively. In Equation (\ref{eq:PD3}), when $y_0'=0$, $\eta(gh(x))$
goes from $1/3$ to 1, from negligible $G$ to the strong field case,
respectively.

It is worth mentioning that when $G$ goes to zero, Equation (\ref{eq:PD3})
is equal to $1/3+4y_0'^2/w'$, and Equation (\ref{eq:RR2D}) for an asymmetric
2D channel reduces to
\begin{equation}
D(x)=\frac{D_0}{\left[1+\frac{1}{4}w'^2(x)\right]^{\frac{1}{3}+4\frac{
y_0'^2}{w'^2}}}.
\label{eq:PD4}
\end{equation}
When $y_0'$ goes to zero the diffusivity for a 2D symmetric channel, as
proposed by RR, is recovered \cite{RR}, and it differs less than 1\%
from the expression obtained by Dagdug and Pineda \cite{DP,Pompa}.

Along with the problem of deriving the modified FJ equation, there are
also questions of the range of applicability of this approximate
one-dimensional description and the accuracy of the expressions for the
effective position-dependent diffusivity. To establish the range of
applicability of the effective diffusivity proposed by Pompa-Garc\'ia
and Dagdug, the present paper focuses on the wide-to-narrow and
narrow-to-wide transitions between the two ends of a conical channel
under a transverse gravitational external field, as shown in Fig.
\ref{tube1}. We study the MFPT for various parameters and observe a
remarkable minimum of the MFPT at intermediate strengths of the
external potential.

\section{Results and discussion}
\label{General results}

We now consider a particle diffusing in a 2D asymmetric conical tube of
length $L$ and variable width $w(x)$ in the presence of the transverse
gravitational external field $G$, as shown in Fig.~\ref{tube1}. Let $\tau
(x_0\rightarrow L)$ denote the particle MFPT from the initial position
given by $x_0$ to the wide end of the channel located at $x=L$, in the
presence of a reflecting boundary at the narrow channel end located at
$x=0$, see the setup in Fig.~\ref{tube1}, panel (a). The MFPT,
considered as a function of $x_0$ and assuming that the reduction to the
effective 1D description is applicable by means of Equation (\ref{FJZwG}),
satisfies \cite{lifsonjackson}
\begin{equation}
\frac{1}{A(x_0)}\frac{d}{dx_0}\left[D(x_0)A(x_0)\frac{d\tau(x_0)}{dx_0}
\right]=-1,
\label{Smmfpt}
\end{equation}
subject to the boundary conditions
\begin{equation}
\tau\eval_{x_0=L}=\frac{d\tau(x_0)}{d x_0}\eval_{x_0=0}=0.
\label{SmmfptBC}
\end{equation}
The solution for $\tau (x_0\rightarrow L)$ is given by
\begin{equation}
\tau(x_0\rightarrow L)=\int_{x_0}^L\frac{dx}{D(x)A(x)}\int_0^xA(y)dy.
\label{SmmfptI}
\end{equation}
The MFPT $\tau_{n\rightarrow w}$ is the MFPT encoded by Equation
(\ref{SmmfptI}) with initial condition $x_0=0$, $\tau_{n\rightarrow w}
=\tau({0\rightarrow L})$. 

Now, let $\tau(x_0\rightarrow 0)$ be the MFPT from the initial position $x_0$
to the narrow end of the channel at $x=0$, in the presence of the reflecting
boundary at the wider channel end at $x=L$, see Fig.~\ref{tube1}, panel
(b). This MFPT satisfies Equation (\ref{Smmfpt}), as well, with the boundary
\begin{equation}
\tau\eval_{x_0=0}=\frac{d\tau(x_0)}{dx_0}\eval_{x_0=L}=0.
\label{SmmfptBC2}
\end{equation}
Integrating Equation (\ref{Smmfpt}) with the boundary conditions
(\ref{SmmfptBC2}) we obtain
\begin{equation}
\tau(x_0\rightarrow0)=\int_0^{x_0}\frac{dx}{D(x)A(x)}\int_x^LA(y)dy.
\label{SmmfptI2}
\end{equation}
The MFPT $\tau_{w\rightarrow n}$ is the MFPT in Equation (\ref{SmmfptI2})
with $x_0=L$, $\tau_{w\rightarrow n}=\tau({L\rightarrow0})$. To compute
the integrals in Equations (\ref{SmmfptI}) and (\ref{SmmfptI2}), $A(x)$
given by Equation (\ref{Asym}), and the interpolation formula for $D(x)$
given by Equations (\ref{eq:RR2D}) and (\ref{eq:PD3}) has to be replaced.

\begin{figure}
\includegraphics[width=0.34\textheight]{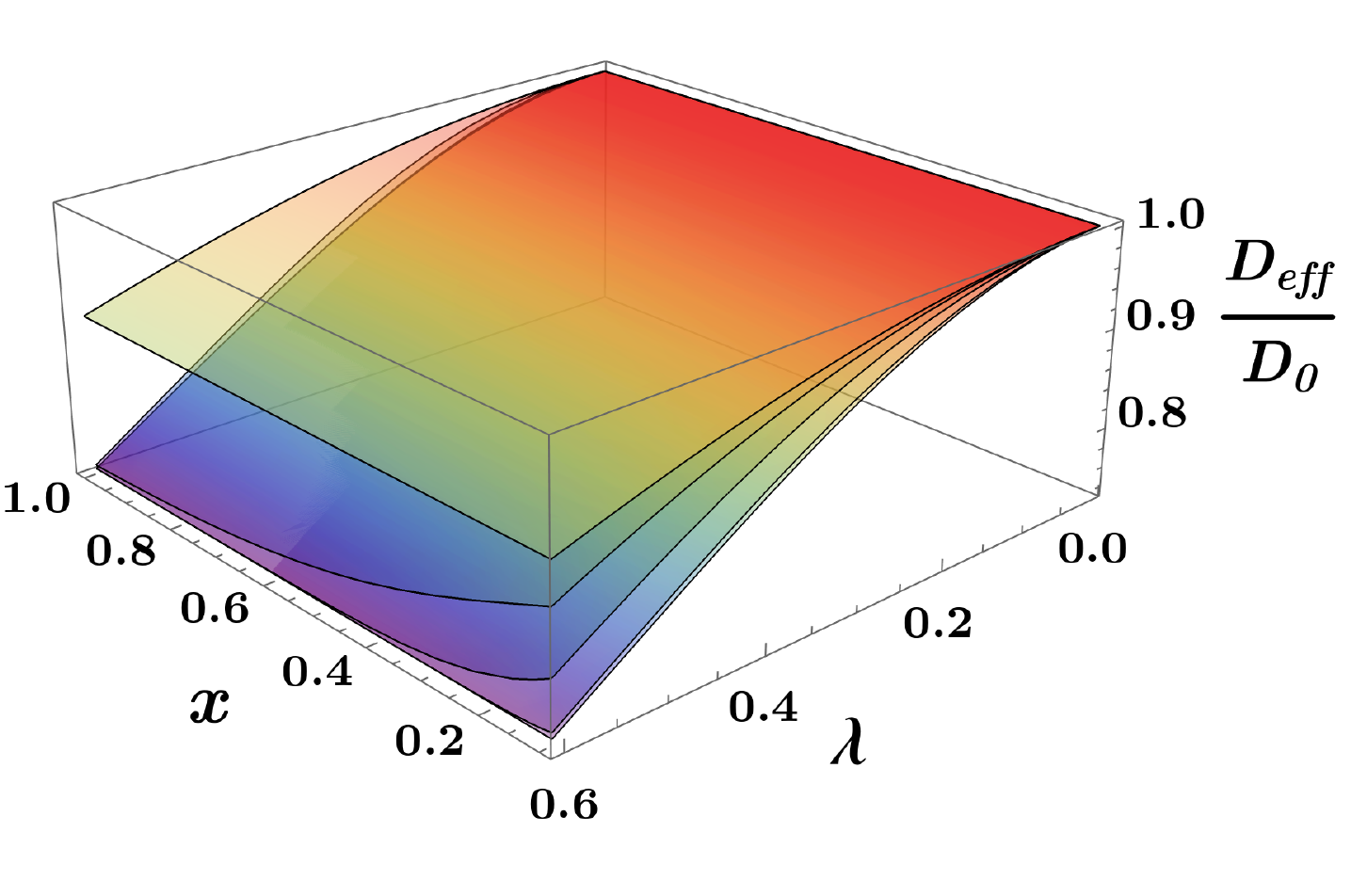}
\includegraphics[width=0.28\textheight]{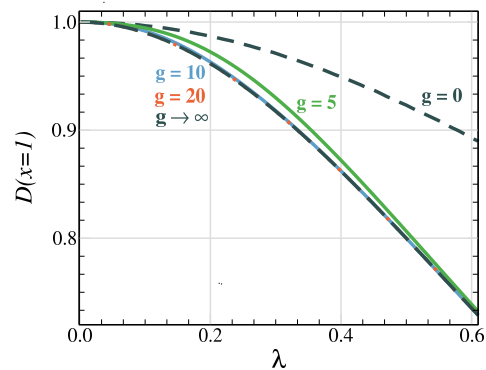}
\caption{Top: Effective diffusion coefficient as function of $x$ and $\lambda$
as predicted by Equation (\ref{eq:RR2D}) and (\ref{eq:PD3}) for a 2D
symmetrical cone-shaped channel formed by straight boundaries $h_1=-\lambda
x-0.1$ and $h_2=\lambda x+0.1$. The channel width variation is given by
$w(x)=h_2(x)-h_1(x)$ and $w'(x)=2\lambda$. The surface graphs from top to
bottom correspond to $g=0$, $5$, $10$, $20$, and $\infty$ (almost
coinciding with the result for $g=20$), respectively.
Bottom: Plot of the effective diffusion coefficient for fixed position
$x=1$ as function of $\lambda$. For $g=10$, $20$, and $\infty$ the curves
are almost indistinguishable.}
\label{Deff}
\end{figure}

In Fig.~\ref{Deff} the effective diffusion coefficient for a conical 2D
channel corresponding to Equations (\ref{eq:RR2D}) and (\ref{eq:PD3}) are
shown for intermediate $g$ values as well as the limiting cases, when $g
\rightarrow0$ and $g\rightarrow\infty$ as function of the position $x$ and
the channel boundary slope $\lambda$. One of the main characteristics of the
surface graphs is that all of them are enclosed between the cases when $g$
tends to zero or infinity. Another important characteristic is that as $x$
increases, the diffusivity for $g$ different from zero tends to the same
value for a fixed value of the channel boundary slope $\lambda$. In the
botton graph of Fig.~\ref{Deff} we show the dependence of the effective
diffusion coefficient at position $x=1$ as function of $\lambda$, for
different values of $g$. The convergence of the behavior for $g\ge10$ is
distinct.

\begin{figure*}
\includegraphics[width=0.46\textwidth]{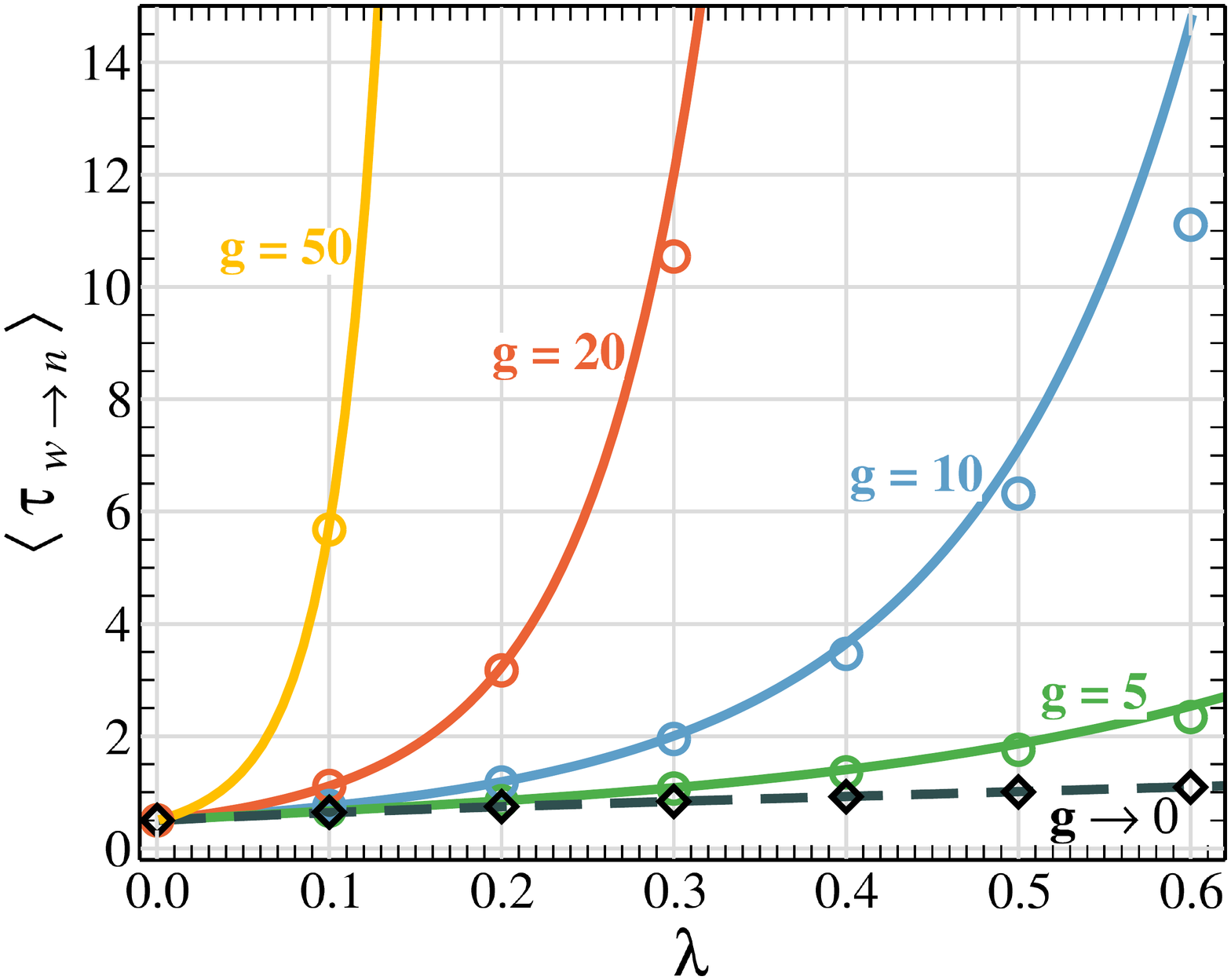}
\includegraphics[width=0.46\textwidth]{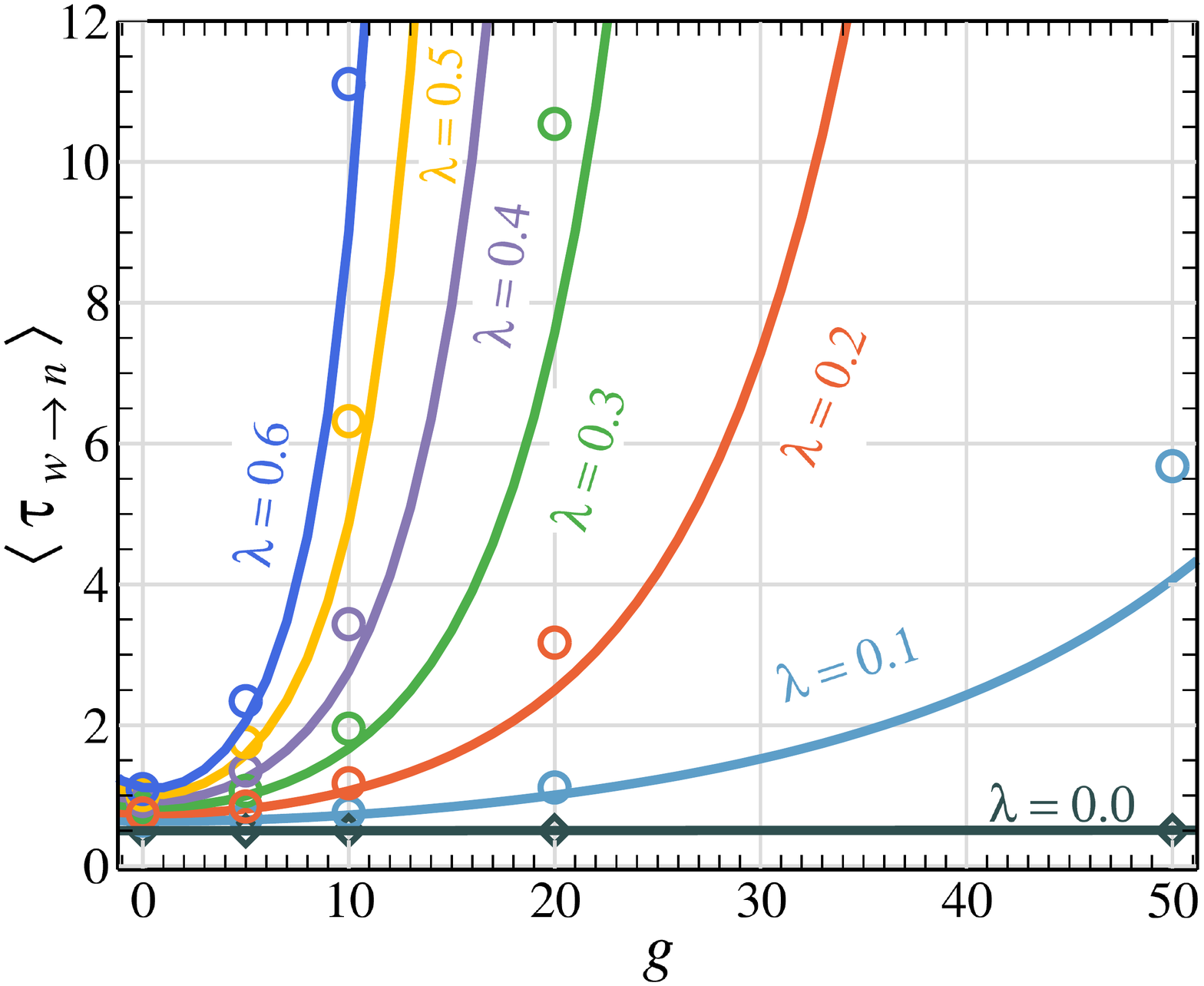}
\caption{MFPT $\tau_{w\rightarrow n}$ for channel passage from the wide to
the narrow end. Left: $\tau_{w \rightarrow n}$ predicted by Equation
(\ref{SmmfptI2}) (continuous lines) compared with the values obtained from
Brownian dynamics simulations (symbols) for different values of force
strength $g$ in a 2D symmetrical cone-shaped channel of length $L=1$. The
channel is formed by the straight and perfectly reflecting boundaries
$h_1(x)=-\lambda x-0.1$, and $h_2(x)=\lambda x+0.1$, see Fig.~\ref{tube1},
panel (b). The limiting case when $g\rightarrow0$ is shown by the dark grey
dashed line. Right: $\tau_{w \rightarrow n}$ predicted by Equation
(\ref{SmmfptI2}) (continuous lines) compared with the values from Brownian
dynamics simulations (symbols) for different values of $\lambda$ as function
of $g$.}
\label{GWN}
\label{GWNxg}
\end{figure*}

Given the expression for the diffusivity and its numerical evaluation, we are
now able to compare the approximate theoretical results for $\tau_{n
\rightarrow w}$ and $\tau_{w\rightarrow n}$ with the corresponding results
from 2D Brownian dynamics simulations. For the purpose of comparison, we
use the same 2D symmetric conical channel from Ref.~\cite{BDB2015} to
perform our simulations. This channel is formed by perfectly reflecting,
symmetric boundaries given by $h_1(x)=-\lambda x-b$ and $h_2(x)=\lambda x
+b$. In the simulations we consider an overdamped, point-like Brownian
particle diffusing inside the 2D conical channel. We describe the particle
dynamics, subject to the constant perpendicular force $G$, by means of the
Langevin equation
\begin{equation}
\frac{d\mathbf{r}}{dt}=\sqrt{2D_0}\boldsymbol{\xi}(t)-G\mathbf{e}_y,
\label{eq:lang}
\end{equation}
where $\mathbf{r}=(x,y)$, and $\boldsymbol{\xi}(t)=(\xi_x(t),\xi_y(t))$ are
zero-mean white Gaussian noise terms with autocorrelation functions $\langle
\xi_i(t),\xi_j(t')\rangle=2\delta_{ij}\delta(t-t')$, where $i,j=x,y$. When
running the simulations we take the time step $\Delta t=10^{-8}$, and the
bulk diffusivity is set to $D_0=1$, so that $\sqrt{2D_0\Delta t}\ll1$.
Finally we set thermal energy to unity, $k_BT=1$. Stochastic averages were
obtained as ensemble averages over $5.0\times10^4$ independent trajectories.

The results for the MFPT $\tau_{w\rightarrow n}$ are shown in Fig.~\ref{GWN}, 
demonstrating very good agreement of the theoretical expressions with the
Brownian dynamics simulations. However, for growing values of the channel
wall slope $\lambda$ we see that the theoretical results overestimates the
MFPT somewhat, especially for the force strength $g=10$ (left panel of
Fig.~\ref{GWN}). In contrast, the MFPT is somewhat underestimated for
intermediate $\lambda$ and larger $g$ values (right panel of Fig.~\ref{GWN}).
In this channel setup, the external force directs the particles towards the
boundary with a positive slope, causing an effective drift away from the
narrow channel end. As can be seen in Fig.~\ref{GWN} the MFPT drastically
increases with growing external force strength. As function of the channel
slope $\lambda$ all curves for different $g$ values converge to a unique
value for $\tau_{w\rightarrow n}$ in the limit of fully horizontal boundaries,
$\lambda=0$. In this case, the channel passage is not affected by the force,
as can also be seen for the effective diffusivity in Fig.~\ref{Deff}. As a
function of the force strength $g$, the MFPT $\tau_{w\rightarrow n}$ has
different values for different channel slopes $\lambda$ in the limit of
vanishing $g$. This behavior is the purely geometric effect of a narrowing
channel. We finally note that both panels show a monotonic increase of the
MFPT $\tau_{w\rightarrow n}$ for growing parameter $g$ as function of the
channel slope $\lambda$ as well as for growing slope $\lambda$ as function
of force strength $g$.

\begin{figure*}
\includegraphics[width=0.46\textwidth]{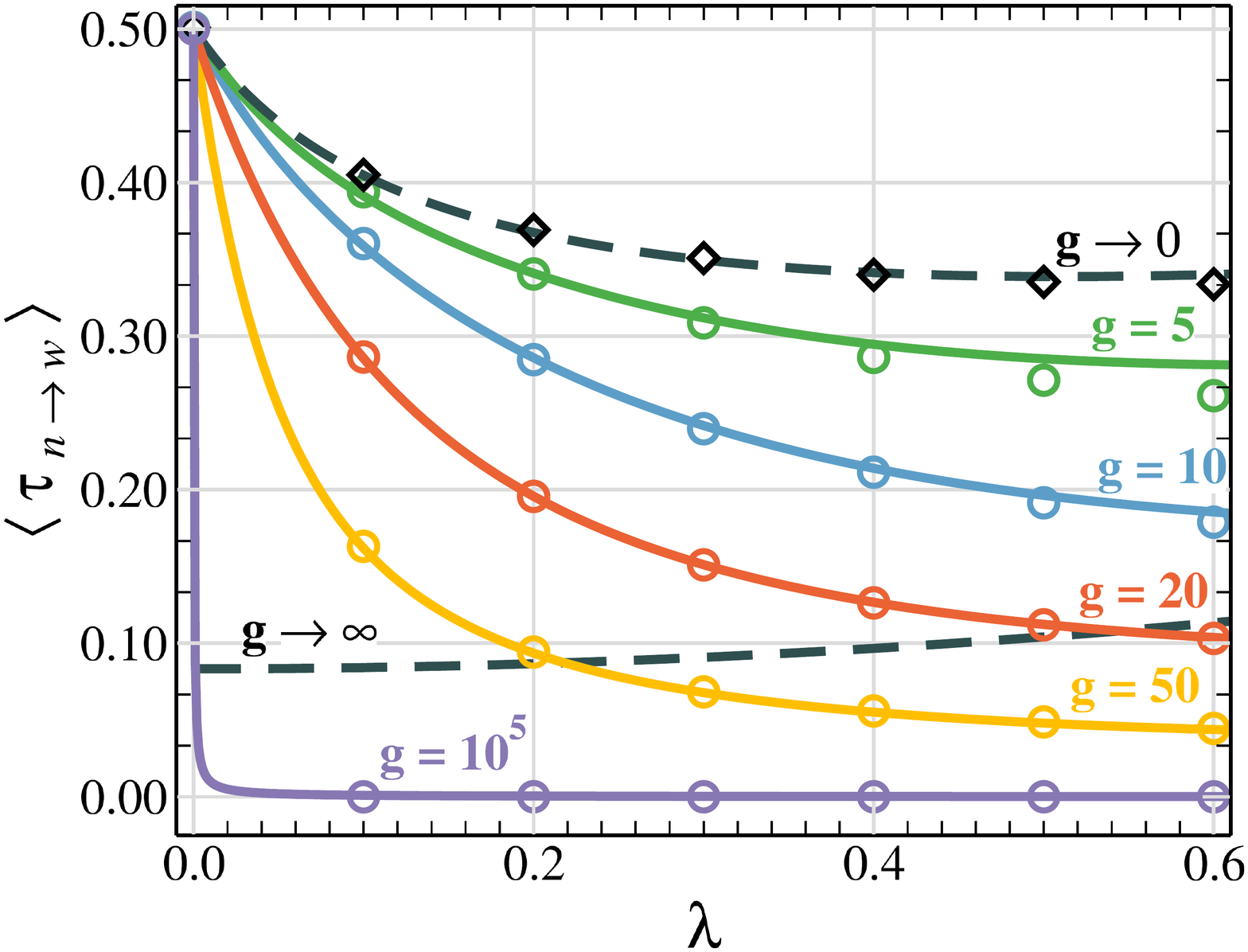}
\includegraphics[width=0.46\textwidth]{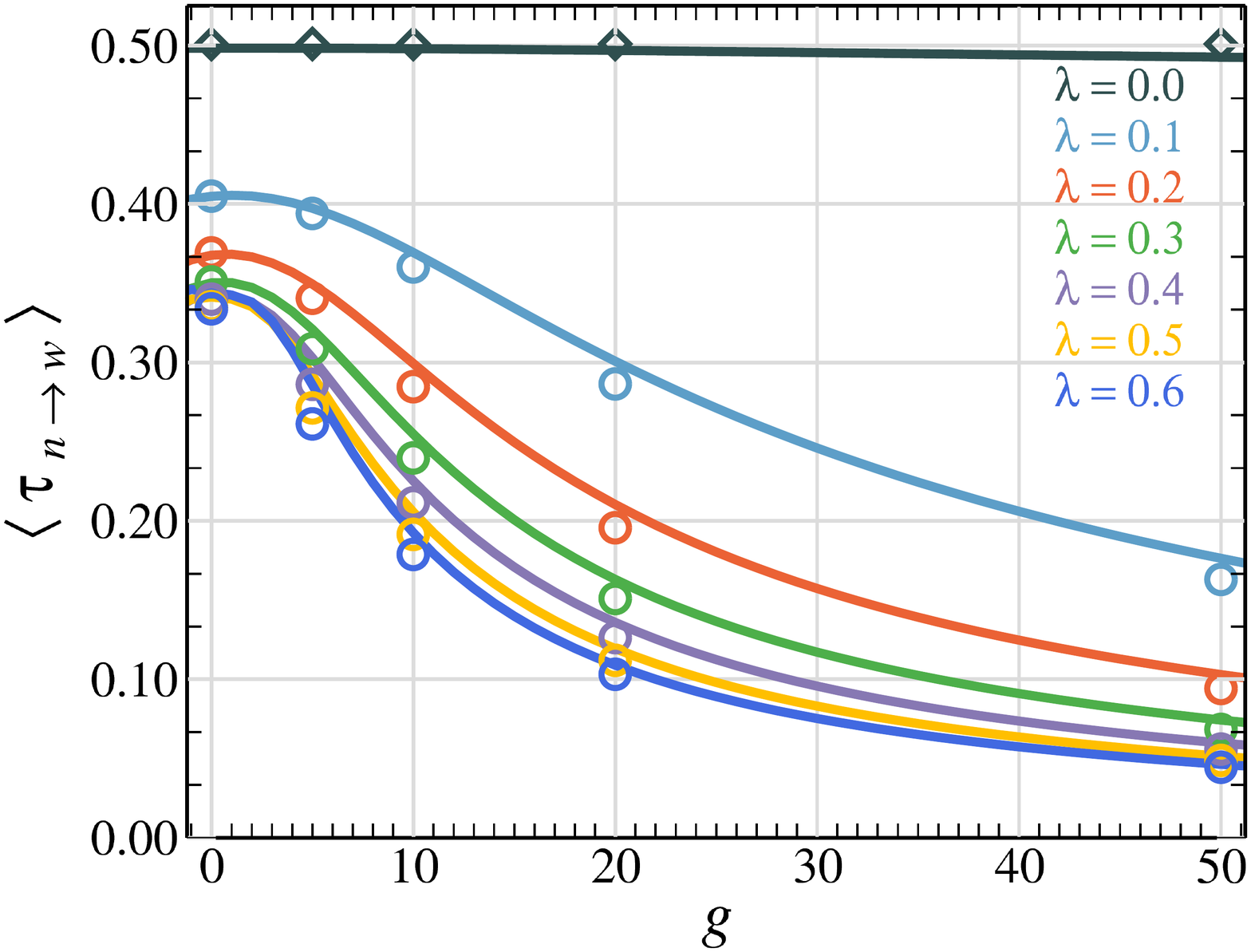}
\caption{MFPT $\tau_{n\rightarrow w}$ for channel passage from the narrow
to the wide end. Left: $\tau_{n\rightarrow w}$ predicted by Equation
(\ref{SmmfptI2}) (continuous lines) compared with the values obtained from
Brownian dynamics simulations (symbols) for different values of force
strength $g$ in a 2D symmetrical cone-shaped channel of length $L=1$. The
channel is formed by the straight and perfectly reflecting boundaries
$h_1(x)=-\lambda x-0.1$ and $h_2(x)=\lambda x+0.1$, see Fig.~\ref{tube1},
panel (a). The limiting cases when $g\rightarrow0$ and $g\to\infty$ are
shown by the dark grey dashed lines. Right: $\tau_{w\rightarrow n}$
predicted by Equation (\ref{SmmfptI2}) (continuous lines) compared with
the values from Brownian dynamics simulations (circles) for different values
of $\lambda$ as a function of $g$.}
\label{GNW}
\label{GNWxg}
\end{figure*}

The results for the MFPT $\tau_{n\rightarrow w}$ for channel passage from
the narrow to the wide end in Fig.~\ref{GNW} also show a very good agreement
between the theoretical predictions and the Brownian dynamics simulations.
Generally, the absolute values for the MFPT are considerably lower than for
the opposite case $\tau_{w\rightarrow n}$. Moreover, the discrepancies are
significantly less for $\tau_{n\rightarrow w}$ for all values of channel
slope $\lambda$ and force strength $g$. Finally, we observe that the value
of $\tau_{n\rightarrow w}$ has the opposite trend as function of $\lambda$
and $g$ as compared to $\tau_{w\rightarrow n}$: here growing slope as well
as increasing force strength lead to an effective drift towards the channel
exit at the wide end, and thus to a reduction of $\tau_{n\rightarrow w}$.
In this narrow-to-wide configuration we observe an interesting result. As
can be seen in the left panel of Fig.~\ref{GNW} there occurs an optimum
for the channel passage at higher values of the channel boundary slope
$\lambda$. Namely, the theoretical result for the MFPT for very high force
strength ($g\to\infty$) exceeds the MFPT values for $g=20$ (the curves
cross at around $\lambda=0.5$) and $g=50$ (crossing at around $\lambda=0.2$).

This crossover behavior with an optimal MFPT warrants some closer inspection,
however. On the one hand it is physically reasonable to argue that such a
minimal MFPT at intermediate $g$ values is the result of the two opposing
effects relevant for higher $g$. Namely, while moderate values of $g$ effect a
resulting drift towards the wide channel exit, when $g$ gets too high it
prevents the particle from exploring the channel in the perpendicular
$y$-direction. The particle therefore cannot profit from the entropic force
pushing it towards the wider channel end. From this observation we can
appreciate the importance of the conspirative interplay between the transverse
external field and the entropic potential. We note that this interplay is hardly
noticed in the right panel of Fig.~\ref{GNWxg} in which the MFPT is depicted as
a function of $g$, for which only moderate $g$-values are shown. On the other
hand, it remains unclear whether the approximations used here to obtain the
effective one-dimensional description with effective entropic forcing remains
valid in the limit $g\to\infty$. Concurrently, we cannot use our computer
simulations to explore the true $g \to \infty$ limit, as the necessary time
steps become prohibitively short with increasing $g$.

From a physical vantage the stark difference in behavior between the plotted
case $g=10^5$ and $g\to\infty$ opens up the possibility of a discontinuous
transition of the MFPT dynamics at $g\to\infty$. Moreover, from a practical
point of view this interesting crossover behavior represents new possibilities
for controlling the transport of Brownian particles in narrow confined
structures for a range of potential applications, including particle
separation, fluid mixing, gating, and catalysis, among others.

\section{Summary and conclusions}
\label{Conclusions}

We studied the crossing dynamics of diffusing point particles of expanding
and narrowing 2D conical channels under the action of a transverse external
gravitational field by means of the MFPT. We derived the theoretical
expression for the MFPT under the assumption that the axial diffusion in the
2D channel with its smoothly varying geometry can be approximately described
as a one-dimensional diffusion in an entropic potential with position-dependent
effective diffusivity in framework of the modified Fick-Jacobs equation. To
this end we use the theoretical expression by Pompa-Garc\'ia and Dagdug
\cite{Pompa} for the interpolation of the effective diffusivity, $D_0/\left[1
+(1/4)w'^2(x)\right]^{-\eta(gw,y'_0)}$, where spatial confinement, asymmetry,
and the presence of a constant transverse force can be encoded in $\eta$, as
a function of the channel width $w$, channel midline $y_0$, and transverse
force $G$ ($g=G/k_BT$). This expression explicitly shows the coupling between
the entropic and energetic effects.

We found very good agreement between the approximate theoretical result for
the MFPT in the two possible configurations: channel passage from the narrow
to the wide end and vice versa. While some deviations are observed for the
wide-to-narrow case at intermediate channel boundary slopes and larger $g$
values, almost perfect agreement is observed for the narrow-to-wide case.
Despite these deviations the general predictions of the approximate 1D
description in terms of the modified Fick-Jacobs equation is validated for
this setting.

A remarkable effect is observed in the narrow-to-wide configuration, where
the theoretical result for the MFPT is not monotonically decreasing with
growing $g$, and thus not bounded by the limiting case $g\to\infty$.
Instead, the MFPT assumes a
minimum at intermediate $g$ values for larger values of the boundary
slope $\lambda$. We interpret this result as an optimum in the interplay
between the effective drift exerted by the entropic potential of the
channel walls (widening towards the channel exit) and a pinning down of
the particle to the channel wall by very high external forcing. In this
case the effect of the entropic force vanishes, and the resulting MFPT
increases. The fact that we can control the exit time of a Brownian particle
from a 2D channel in the presence of an external transverse force may allow
the development of practical applications including particle separation,
gating, controlling effective fluid mixing, and catalysis, among others.
However, even though the theoretical expressions can be used to predict this
crossover behavior of the MFPT values under the influence of large external
forces, their full range of applicability remains somewhat tricky to establish.
The theoretical affirmation is based on the fact that we are trying to predict
the behavior of a two-dimensional system while using an effective diffusivity
obtained by means of a dimensional reduction, that removes the degree of
freedom, which coincides with the direction of the applied force. Moreover,
the effective diffusivity model, and as a consecuence, the expressions for
the MFPT, does not contain any information about the direct interaction
between the boundary walls and the particle further than its role as a
boundary condition. Concurrently, while for moderate to large $g$ values the
simulations tend to be close to numerical results, simulations
of confined particles under a very high potential is subtle to implement
because the conditions of the system require a small time step and an
appropriate particle-wall interaction, involving a high computational
cost. Thus, while this effect is physically interesting and, to our
knowledge, reported for the first time, further research is needed to
exactly establish the precise quantitative behavior.

It will be interesting to verify the effect of an external forcing
perpendicular to the symmetry axis of the channel in 3D settings as well as
for channels filled with complex liquids, e.g., when the particle exhibits
viscoelastic subdiffusion.

\begin{acknowledgments}
This study was partially supported by CONACyT under the grant Frontiers
Science No. 51476. Financial support from German Science Foundation (DFG,
grant no. ME 1535/12-1) is acknowledged.
\end{acknowledgments}

\textbf{Data availability statement.}  The data that support the
findings of this study are available from the corresponding author upon
reasonable request.

\end{document}